\documentclass[twocolumn,showpacs,preprintnumbers,floatfix]{revtex4}
\usepackage{amssymb,amsmath,amsfonts}
\usepackage{graphicx} 
\usepackage{bm} 

\begin{document}
\title{Modular organization as a basis for the functional integration/segregation in large-scale brain networks }
\author{M. Valencia,$^{1}$ M.~A. Pastor,$^{2}$ MA. Fern{\'a}ndez-Seara,$^{2}$ J. Artieda,$^{2}$
J. Martinerie,$^{1}$ and M. Chavez$^{1}$}
\affiliation{%
$^1$~Laboratoire de Neurosciences Cognitives et Imagerie
C\'{e}r\'{e}brale. LENA-CNRS UPR-640, Paris, France}%
\affiliation{%
$^2$~Department of the Neurological Sciences, Center for Applied
Medical Research, University of Navarra School of Medicine and
Cl{\'i}nica Universitaria de Navarra, Pamplona, Spain}
%
\begin{abstract}
Modular structure is ubiquitous among real-world networks from related proteins to social groups.  Here we analyze the modular organization of brain networks at a large-scale (voxel level) extracted from functional magnetic resonance imaging (fMRI) signals. By using a random walk-based method, we unveil the modularity of brain-webs, and show modules with a spatial distribution that matches anatomical structures with functional significance.  The functional role of each node in the network is studied by analyzing its  patterns of inter- and intra-modular connections. Results suggest that the modular architecture constitutes the structural basis for the coexistence of  functional integration of distant and specialized brain areas during normal brain activities at rest. 
\end{abstract}


\maketitle %

\textbf{There is a growing interest in studying the connectivity patterns extracted from brain signals during different mental states. Current studies suggest that brain architecture leads neural assemblies to be coordinated with an optimized wiring cost. Brain webs coordinate a mosaic of brain modules,  carrying out specific functional tasks and integrated into a coherent process. We analyze the modular structure of brain networks extracted from fMRI signals in humans at rest. Using a random walk-based method we identify a non-random modular architecture of brain connectivity.  This approach is fully data driven and relies on no a priori choice of a seed brain region or signal averaging in predefined brain areas. The analysis of intra- and inter-modules connections leads us to relate a node's connectivity  to a local information processing, or to the integration of distant anatomo/functional brain regions. We also find that the spatial distribution of the retrieved modules matches with brain areas associated with specific functions, assessing a functional significance to the modules. In our conclusions, we argue that a modular characterization of the functional brain webs constitutes an interesting model for the study of brain connectivity during different pathological or cognitive states.}

\section{Introduction}
From the brain over the Internet to social groups, complex networks are a prominent framework to describe collective behaviors in many areas~\cite{Boccaletti2006}.  Many of real-world networks exhibit topological features that can be captured neither by regular connectivity models as lattices, nor by random configurations~\cite{Watts1998, Barabasi1999}.  Under this framework, recent studies of complex brain networks have attempted to characterize the connectivity patterns observed under functional brain states. Electroencephalography (EEG), magnetoencephalography (MEG), or functional magnetic resonance imaging (fMRI) studies have consistently shown that human brain functional networks during different pathological and cognitive neurodynamical states display small world (SW) attributes~\cite{Salvador2005a, Eguiluz2005, Achard2006, Achard2007, Bassett2006, Reijneveld07}.  SW networks are characterized by a small average distance between any two nodes while keeping a relatively highly clustered structure. Thus, SW architecture is an attractive model for brain connectivity because it leads distributed neural assemblies to be integrated into a coherent process with an optimized wiring cost~\cite{LagoFernandez00, buzsaki04}.

Another property observed in many networks is the existence of a modular organization in the wiring structure. Examples range from RNA structures to biological organisms and social groups. A module is currently defined as a subset of units within a network such that connections between them are denser than connections with the rest of the network. It is generally acknowledged that modularity increases robustness, flexibility and stability of biological systems~\cite{Barabasi2004,Sole2008}. The widespread character of modular architecture in real-world networks suggests that a network's function is strongly ruled by the organization of their structural subgroups.

Empirical studies have lead to the hypothesis that specialized neural populations are largely distributed and linked to form a web-like structure~\cite{Varela2001}.  The emergence of any unified brain process relies on the coordination of a scattered mosaic of modules, representing functional units, separable from -but related to- other modules. Characterizing the modular structure of the brain may be crucial to understand its organization during different pathological or cognitive states.

Previous studies over the mammalian and human brain networks have successfully used different methods to identify clusters of brain activities. Some classical approaches, such as those based on principal components analysis (PCA) and independent components analysis (ICA), make very strong statistical assumptions (orthogonality and statistical independence of the retrieved components, respectively) with no physiological justification~\cite{Biswal1999, McKeown2003}. Although a number of studies investigating the organization of anatomic and functional brain networks have shown very interesting properties of the macro-scale brain architecture~\cite{Hagmann2008, Chen2008}, little is known about the network structure at a finer scale (at a voxel level).  Current approaches are based on the use of a priori coarse parcellations of the cortex~\cite{Salvador2005a, Achard2006}; or on partial networks defined by a seed voxel~\cite{Cordes2001}. Nevertheless, seed-based descriptions may fail to describe the global behavior of the brain, as they only consider the connectivity of the reference voxel. On the other hand, parcellation schemes reduce the analysis to a macro-scale fixed by an \emph{a priori} definition of the brain areas. Further, a recent study shows that the topological organization of brain networks is affected by the different parcellation strategies applied \cite{Wang2008}.

Here we focus on a completely data-driven framework to study the connectivity of brain networks extracted directly from functional magnetic resonance imaging (fMRI) signals at voxel resolution. A random walk-based algorithm is used to assess the modular organization of functional networks from healthy subjects in a resting-state condition. Results reveal that functional brain webs present a large-scale modular organization significatively different from that arising from random configurations. 
Further, the spatial distribution of some modules fits well with previously defined anatomo-functional brain areas, assessing a functional significance to the retrieved modules. Based on the patterns of inter- and intra-modular connectivities, we also study the  roles played by different brain sites~\cite{Guimera2005}.  Results provide a characterization of the functional scaffold that underly the coordination of specialized brain systems during spontaneous brain behavior.
  
\section{Data adquisition and preprocessing}
BOLD fMRI data were acquired using a T2*-weighted imaging sequence during a period of 10 minutes from 7 healthy right-handed subjects. The study was performed with written consent of the subjects and with the approval of local ethics committees.  During the scan, all subjects were instructed to rest quietly, but alert, and keep their eyes closed. 500 volumes of gradient echoplanar imaging (EPI) data depicting BOLD contrast were acquired. In the acquisition, we used the following parameters: number of slices, $21$ (interleaved); slice thickness, $4$ mm; inter-slice gap, $1$ mm; matrix size, $64 \times 64$; flip angle, $90\,^{\circ}$; repetition time (TR), $1250$ ms; echo time, $30$ ms; in-plane resolution, $3 \times 3$ mm$^{2}$. Subsequently, a high resolution structural volume was acquired via a T1--weighted sequence (axial; matrix $192\times256\times160$; FOV $192\times256\times160$ mm$^{3}$; slice thickness; $1$ mm; in--plane voxel size, $1\times1$ mm$^{2}$; flip angle 15$\,^{\circ}$ ; TR, $1620$ ms, TI, $950$ ms; TE, 3.87 ms)  to provide the anatomical reference for the functional scan. 

All acquired brain volumes were corrected for motion and differences in slice acquisition times using the SPM5 (http://www.fil.ion.ucl.ac.uk) software package. After correction, fMRI datasets were coregistered to the anatomical dataset and normalized to the standard template MNI, enabling comparisons between subjects. Due to computational limitations, normalized and corrected functional scans were subsampled to a 4x4x4~mm resolution, yielding a total of 20898 voxels (nodes in the network). To eliminate low frequency noise (e.g.~slow scanner drifts) and higher frequency artifacts from cardiac and respiratory oscillations, time-series were digitally filtered with a finite impulse response (FIR) filter with zero-phase distortion (bandwidth $0.01-0.1$~Hz)~\cite{Cordes2001}. 

\section{Estimation of functional connectivity}
A functional link between two time series $x_i(t)$ and $x_j(t)$ (normalized to zero mean and unit variance) was defined by means of the linear cross-correlation coefficient computed as $r_{ij} = \langle x_i(t)x_j(t)\rangle $, where $\langle\cdot\rangle$ denotes the temporal average. For the sake of simplicity, we only considered here correlations at lag zero. To determine the probability that correlation values are significantly higher than what is expected from independent time series, $r_{ij}(0)$ values (denoted $r_{ij}$) were firstly transformed by the Fisher's Z transform 
\begin{equation}
Z_{ij} = 0.5\ln \left(\frac{1+r_{ij}}{1-r_{ij}} \right)
\end{equation}
Under the hypothesis of independence, $Z_{ij}$ has a normal distribution with expected value 0 and variance $1/(df-3)$, where $df$ is the effective number of degrees of freedom~\cite{Bartlett1946,Bayley1946,Jenkins1968}. If time series are formed of independent measurements, $df$ simply equals the sample size, $N$.  Nevertheless,  autocorrelated time series do not meet the assumption of independence required by the standard significance test, yielding a greater Type I error~\cite{Bartlett1946,Bayley1946,Jenkins1968}. In presence of auto-correlated time series $df$ must be corrected by the following approximation: 
\begin{equation}
\frac{1}{df}\approx \frac{1}{N} + \frac{2}{N}\sum_\tau r_{ii}(\tau) r_{jj}(\tau), 
\end{equation}
where $r_{xx}(\tau)$ is the autocorrelation of signal $x$ at lag $\tau$. Other estimators of $df$, and statistical significance tests for auto-correlated time series can be found in~\cite{Pyper1998}. To correct for multiple testing, the False Discovery Rate (FDR) method was applied to each matrix of $r_{ij}$ values~\cite{Benjamini1995}. With this approach, the threshold of significance $r_{\text{th}}$ was set such that the expected fraction of false positives is restricted to $q \leq 0.001$.

\begin{figure}[!htb]
   \centering
   \resizebox{0.5\columnwidth}{!}{\includegraphics{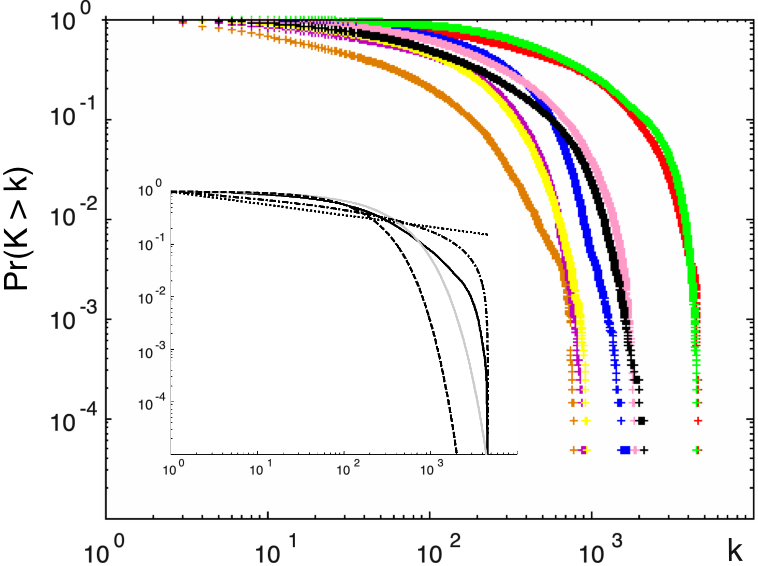}}
   \caption{Cumulative degree distributions $P(K>k)$ estimated from all subjects.
   The inset depicts the average degree distribution. Black and gray
   solid lines correspond to the observed distribution and the fitted truncated power law model, respectively. Dotted, dashed and dot-dash lines correspond to fitted power law, exponential law and truncated Pareto law distributions, respectively.}\label{netDegreeCdf}
\end{figure}

In the construction of the networks, a functional connection between two brain sites was assumed as an undirected and unweighted edge ($A_{ij} = 1$ if $r_{ij} > r_{\text{th}}$; and zero otherwise). Although topological features can also be straightforwardly generalized to weighted networks, we obtained qualitative similar results (not reported here) for weighted networks with a functional connectivity strength between nodes given by $w_{ij} = r_{ij}$.

To characterize the topological properties of a network, a number of parameters have been described. Here we use three key parameters: mean degree $\langle K \rangle$, clustering index $C$ and global efficiency $E$~\cite{Watts1998, Barabasi1999, Boccaletti2006}. Briefly, the degree $k_{i}$ of node $i$ denotes the number of functional links incident with the node and the mean degree is obtained by averaging $k_{i}$ across all nodes of the network. The clustering index quantifies the local density of connections in a node's neighborhood. For a node $i$, the clustering coefficient $c_{i}$ is calculated as the number of links between the node's neighbors divided by all of their possible connections and $C$ is defined as the average of $c_{i}$ taken over all nodes of the network. The global efficiency $E$ provides a measure of the network's capability for information transfer between nodes and is defined as the inverse of the harmonic mean of the shortest path length $L_{ij}$ between each pair of nodes. 

Figure~\ref{netDegreeCdf} shows superimposed the degree distributions for the seven studied subjects. For each network, goodness-of-fit was compared here using Maximum Likelihood methods and the Kolmogorov-Smirnov statistic (KS) for four possible forms of degree distribution $p(k)$: a power law $p(k) \propto k^{-\gamma}$; an exponential $p(k) \propto \exp ^{-\lambda k}$; a truncated Pareto $p(k) \propto (\nu ^{\alpha+1} - \zeta^{\alpha+1})^{-1} k ^{\alpha}$; and an exponentially truncated power law $p(k) \propto k ^{\alpha-1} \exp(-k/k_c)$. The bestfitting were obtained for the truncated power law ($\text{KS} = 0.0421$ compared with $\text{KS} = 0.1028$, $0.2632$ and $0.3278$ for the exponential law, the truncated Pareto and the power law distribution, respectively). Estimated parameters for the truncated power law are $\alpha  = 0.7688 \pm 0.1455$, $k_{c} = 410 \pm 351$. 

\begin{table}
\begin{center}
\begin{tabular}{l|ccccccc}
$\text{S}_i$ & S1 & S2 & S3 & S4 & S5 & S6 & S7 \\
\hline
$\langle K\rangle$ &  
710.65	&
248.37	&
815.17	&
263.59	&
134.69	&
133.06	&
201.16	\\
$C$ $^*$& 
0.4954	&
0.3901	&
0.4865	&
0.3856	&
0.3541	&
0.3389	&
0.3638	\\
$\overline{C}_{rnd}$ &  
0.0340	&
0.0119	&
0.0390	&
0.0126	&
0.0064	&
0.0064	&
0.0096	\\
$E$ & 
0.3888	&
0.3569	&
0.4135	&
0.3447	&
0.3104	&
0.3132	&
0.3269	\\
$\overline{E}_{rnd}$ &
0.5170	&
0.4973	&
0.5195	&
0.5004	&
0.4337	&
0.4322	&
0.4810	
\end{tabular}
\end{center}
\caption{Parameters for real and randomized networks: $\langle K\rangle$, mean degree; $C$, clustering index; $E$, global efficiency; $\overline{\theta}_{rnd}$ denotes the average of parameter $\theta$ obtained from $10$ equivalent randomized networks. Single asterisks indicate that this parameter has a significance level of  $p<10^{-4}$.}
\label{tableForNetsParams}
\end{table}

Values of the topological parameters are summarized in Table~\ref{tableForNetsParams}.  To asses the statistical significance of brain connectivity, we perform a benchmark comparison of the functional connectivity patterns. For this, the topological features of brain webs are compared with those obtained from equivalent random wirings.  To create an ensemble of equivalent random networks we use the algorithm described in ~\cite{Boccaletti2006}. According to this procedure, each edge of the original network is randomly rewired avoiding self- and duplicate connections. The obtained randomized networks thus preserve the same mean degree as the original network, whereas the rest of the wiring structure is random. The significance of a given topological parameter $\theta$ was assessed by quantifying its statistical deviation from values obtained for the ensemble of randomized networks. Let $\mu$ and $\sigma$ be the mean and standard deviation of the parameter $\theta$ computed from such an ensemble. The significance is given by the ratio $\Sigma=(\theta - \mu)/\sigma$ whose p-value is given by the Chebyshev's inequality (for \emph{any} statistical distribution of $\theta $: $p(|\Sigma| \geqslant \zeta) \leqslant 1/\zeta^{2}$ where $\zeta$ is the chosen statistical threshold)~\cite{Papoulis1991}. 

The sparse connectivity of functional brain networks was found to be significatively different from randomized wirings for all the subjects. Brain networks yielded larger clustering values ($p<10^{-4}$) than the equivalent random configurations, but similar efficiency values, indicating a connectivity with SW attributes. 

\section{Modular analysis of brain networks} 
A potential modularity of brain-webs is suggested by the fact that brain networks display a clustering index approximately one order of magnitude larger than that obtained from random configurations~\cite{Ravasz2002}. Although the notion of module results very intuitive, in general it is difficult to define formally. It is currently accepted that a partition $\mathcal{P} = \{ \mathcal{C}_{1},\ldots,\mathcal{C}_{M} \}$ represents a good modular structure if the portion of edges inside each module $\mathcal{C}_{i}$ (intra--modular edges) is high compared to the portion of edges between them (inter--modular edges).  The modularity $Q(\mathcal{P})$, for a given partition $\mathcal{P}$ of a network
is formally defined as~\cite{Newman2004}: 
\begin{equation}
Q(\mathcal{P})=\sum_{s=1}^{M}{\left[{\frac{l_{s}}{L}-\left(\frac{k_{s}}{2L}\right)}^{2}\right]},
\end{equation} 
where $M$ is the number of modules, $L$ is the total number of connections in the network, $l_{s}$ is the number of connections between vertices in module $s$, and $k_{s}$ is the sum of the degrees of the vertices in module $s$.

\begin{table}[!htb]
\begin{center}
\begin{tabular}{l|ccccccc} 
& S1 & S2 & S3 & S4 & S5 & S6 & S7 \\
\hline
$Q^{**}$ & 0.4385 & 0.5814 & 0.4223 & 0.5538 & 0.5648 & 0.5749 & 0.5362\\ 
$\overline{Q}_{rnd}$& 0.0065 & 0.0169 & 0.0057 & 0.0160 & 0.0274 & 0.0279  & 0.0201 \\ 
\end{tabular}
\end{center}
\caption{Modularity for real ($Q$) and randomized networks. $\overline{Q}_{rnd}$ denotes the average obtained from $10$ equivalent randomized networks. Double asterisks denotes a significance level of  $p<10^{-6}$.}
\label{tableForNetsModularity}
\end{table}

 \begin{figure*}[!htb]
   \centering
   \resizebox{\textwidth}{!}{\includegraphics{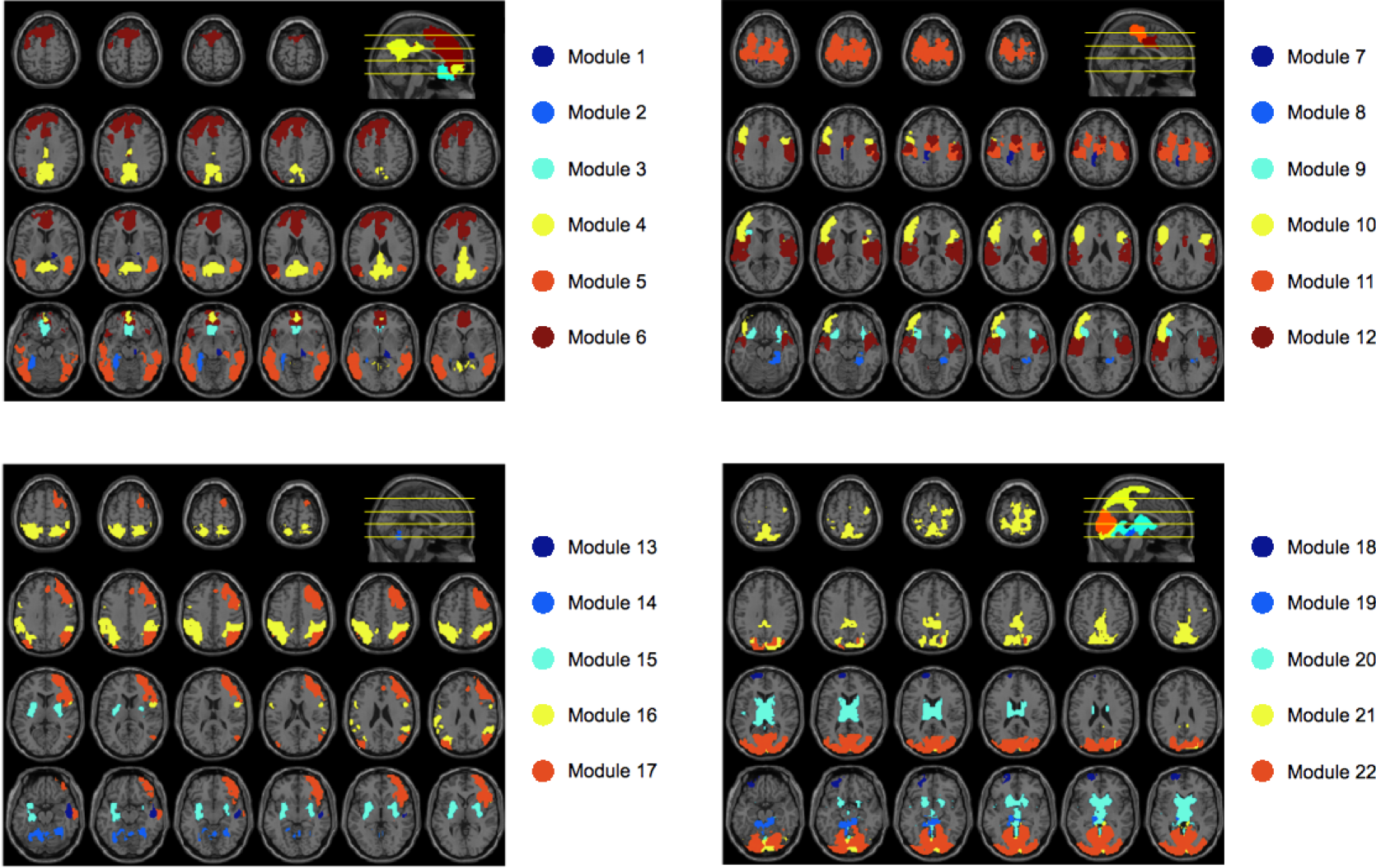}}
   \caption{Main functional brain modules: brain sites belonging to each module were coloured and superimposed onto an anatomical image. The sagittal anatomical images at the top right of each plot indicate the relative position of imaged slices of each row. For the sake of clarity, we show only those communities with a size larger than 40 voxels ($\sim 0.2\%$ of the size of the whole network).}
\label{commsDistribAndNet}
 \end{figure*}

To partition the functional networks in modules, we used a random walk-based algorithm~\cite{Pons2006}, because of its ability to manage very large networks, and its good performances in benchmark tests~\cite{Pons2006, Danon2005}. In a nutshell, a random walker on a connected graph tends to remain into densely connected subsets corresponding to modules. Let $P_{ij}= \frac{A_{ij}}{k_{i}}$ to be the transition probability from node $i$ to node $j$, where $A_{ij}$ denotes the adjacency matrix and $k_{i}$ is the degree of the i$^{\text{th}}$ node. This defines the transition matrix $(P^t)_{ij}$ for a random walk process of length $t$ (denoted here $P^t_{ij}$ for simplicity). The metric used to quantify the structural similarity between vertices is given by 
\begin{equation}
\rho_{ij} = \sqrt{\sum_{l=1}^{N}\frac{(P^{t}_{il}-P^{t}_{jl})^{2}}{k_{l}}}
\end{equation}
Using matrix identities, the distance $\rho$ can be written as  $\rho^{2}_{ij}=\sum^{n}_{\alpha=2}{\lambda^{2t}_{\alpha}{(v_{\alpha}(i)-v_{\alpha}(j))}^{2}}$; where $(\lambda_{\alpha})_{1   \leqslant \alpha  \leqslant n}$ and $(v_{\alpha})_{1 \leqslant \alpha \leqslant n}$ are the $n$ eigenvalues and right eigenvectors of the matrix $P$, respectively~\cite{Pons2006}. This relates the random walk algorithm to current methods using spectral properties of the graphs~\cite{Newman2006,Gfeller2007}. The current approach, however, needs not to explicitly compute the eigenvectors of the matrix; a computation that rapidly becomes intractable when the size of the graphs exceeds some thousands of vertices.

To find the modular structure, the algorithm starts with a partition in which each node in the network is the sole member of a module. Modules are then merged by an agglomerative approach based on a hierarchical clustering method. Following Ref.~\cite{Pons2006}, if two modules $\mathcal{C}_{1}$ and $\mathcal{C}_{2}$ are merged into a new one $\mathcal{C}_{3} = \mathcal{C}_{1} \cup \mathcal{C}_{2}$, the transition matrix is updated as follows: $P^t_{\mathcal{C}_3 k}=\frac{|\mathcal{C}_1|P^t_{\mathcal{C}_1 k}+|\mathcal{C}_2|P^t_{\mathcal{C}_2 k}}{|\mathcal{C}_1|+|\mathcal{C}_2|}$, where $|\mathcal{C}_{i}|$ denotes the number of elements in module $\mathcal{C}_{i}$. The algorithm stops when all the nodes are grouped into a single component. At each step the algorithm evaluates the quality of partition $Q$. The partition that maximizes $Q$ is considered as the partition that better captures the modular structure of the network. In the calculation of $Q$, the algorithm excludes small isolated groups of connected vertices without any links to the main network. However, these isolated modules are considered here as part of the network for the calculation of the topological parameters.

As reported in Table~\ref{tableForNetsModularity}, a modular structure is confirmed by the high values of $Q$ obtained for the optimal partition of the networks (a value of $Q \geq 0.3$ is in practice a good indicator of modularity in a network~\cite{Clauset2004}). Further, values of modularity for all the subjects were statistically significant when compared with randomized wirings ($p<10^{-6}$). To assess the stability of the partition structure across subjects  we used the Rand index $J$~\cite{Rand1971}, which is a traditional criterion for comparison of different results provided by classifiers and clustering algorithms, including partitions with different numbers of classes or clusters. For two partitions $P$ and $P'$ the Rand index is defined as $J=\frac{a+d}{a+b+c+d}$; where $a$ is number of pairs of data objects belonging to the same class in $P$ and to the same class in $P'$, $b$ is number of pairs of data objects belonging to the same class in $P$ and to different classes in $P'$, $c$ is the number of pairs of data objects belonging to different classes in $P$ and to the same class in $P'$, and $d$ is number of pairs of data objects belonging to different classes in $P$ and to different classes in $P'$. Thus index $J$ yields a normalized value between $0$ (if the two partitions are randomly drawn) and $1$ (for identical partition structures).  For our data, the values of $J$ indicate a moderate stability of the partition structure across all subjects ($J = 0.5148 $).  

To assess a functionality to the different groups of the modular brain webs, we compared the spatial distribution of the recovered modules with a previously reported  anatomical parcellation of the human brain~\cite{Tzourio-Mazoyer2002}. For the sake of simplicity, we only consider here communities whose size was larger than 40 voxels (${\sim0.2\%}$ of the size of the whole network), which yields $N_C = 22$ modules. 

Fig.~\ref{commsDistribAndNet} illustrates the spatial distribution of the modules retrieved from the averaged connectivity matrix computed over all subjects. Results show that the spatial distribution of recovered modules fits well some brain systems. Module 22 for instance, includes $75\%$ of the primary visual areas V1, while module 5 overlaps half of the ventral visual stream (brain areas V2 and V4), and visual areas of the V3 region (cuneus and precuneus)  are included ($\sim40\%$) in the module 4. Module 20 includes most of the subcortical structures caudate  and thalamus nuclei  (covered at $70\%$ and $75\%$, respectively). The auditory system is included by module 12 that overlaps primary and secondary areas plus associative auditory cortex ($60-70\%$).   Modules 11, 16 and 21 cover most ($40-70\%$) of the somatosensory and motor cortices; and language related areas are mainly included  ($>60\%$) in module 10. 

Importantly, some modules include distant brain locations that are functionally related, e.g. the language related areas (modules 10), the auditory system (module 12), or brain regions involved in high level visual processing tasks (module 5). This spatially distributed organization of modules rules out the possibility that modularity \emph{simply} emerges as a consequence of vascular processes or local physiological activities independent of neuronal functions \cite{Logothetis2001, Fox2007}. 
 
Modules assignment provides the basis for the classification of nodes according to their patterns of intra- and inter-modules connections, which conveys significant information about the importance of each node within the network~\cite{Guimera2005}. 

The within-module degree $z$-score measures how well connected the node $i$ is to other nodes in the module, and is defined as:
\begin{equation}
z_{i} = \frac{k_{i}-\overline{k}_{s_{i}}}{\sigma_{ks_{i}}}
\end{equation}
where $k_{i}$ is the number of links of node $i$ to other nodes in its module $s_{i}$, $\overline{k}_{s_{i}}$ is the average of $k$ over all the nodes in $s_{i}$, and $\sigma_{k_{s_{i}}}$ is the standard deviation of $k$ in $s_{i}$.  Thus node $i$ will display a large value of  $z_{i}$ if it has a large number of intra-modular connections relative to other nodes in the same module, i.e. it measures how well connected a node is to other nodes in the module).

The extent a node $i$ connects to different modules is measured by the participation coefficient $pc_{i}$ defined as:
 \begin{equation}
pc_{i} = 1-\sum_{s=1}^{M} \left( \frac{k_{is}}{k_{i}}\right)^{2}
\end{equation}
 where $k_{is}$ is the number of links of node $i$ to nodes in module $s$, and $k_{i}$ is the degree of node $i$.  The participation coefficient takes values of zero if a node has most of its connections exclusively with other nodes of its module. In contrast, $pc_{i} \sim 1$  if their links are distributed among different modules in the network.

\begin{figure}
   \centering
   \resizebox{0.75\columnwidth}{!}{\includegraphics{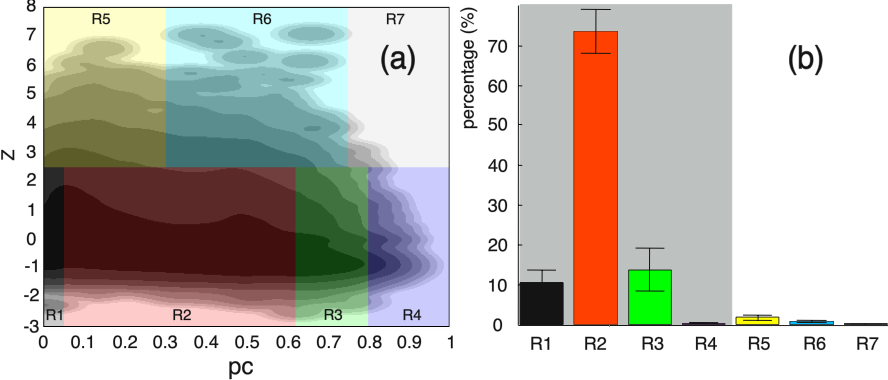}}
   \caption{Role determination, as represented in the $z-pc$ parameter space. (a) Average density landscape (computed over all networks) depicted in logarithmic scale. (b) Histograms and error bars corresponding to the proportion of nodes for each role. Histograms are coloured according to the roles depicted in the $z-pc$ parameter space.}
    \label{rolesBrainNets}
 \end{figure}

\begin{figure*}[!htbp]
   \centering
   \resizebox{\textwidth}{!}{\includegraphics{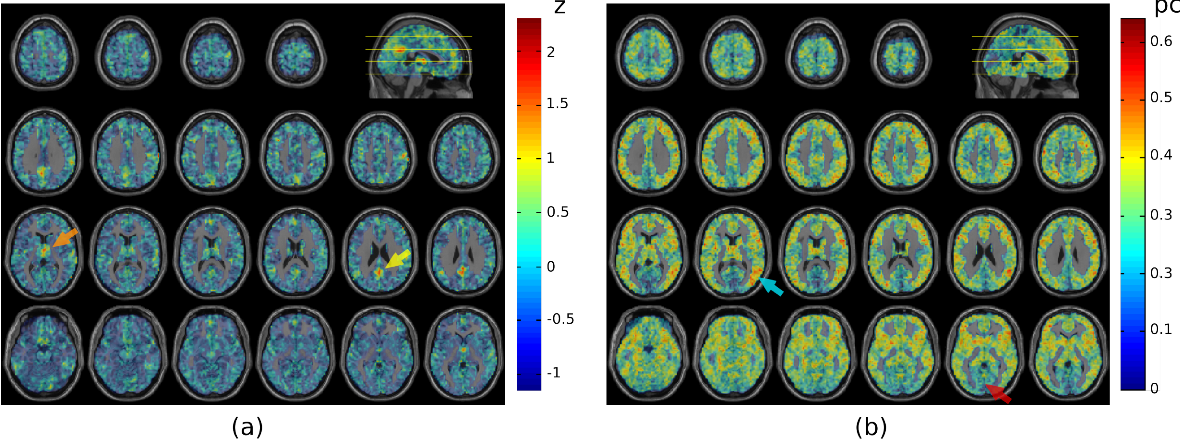}}
   \caption{Anatomical distribution of  (left) the averaged within-module degree $z$, and (right) the averaged participation coefficient $pc$ indices computed across all subjects. The sagittal anatomical images at the top right of each plot indicate the relative position of imaged slices of each row.}
    \label{rolesBrainDistrib}
 \end{figure*}

The role (R$_{i}$) of a node in the network can be assessed by its within-module degree and its participation coefficient, which define how the node is positioned in its own module and with respect to other modules~\cite{Guimera2005}.  Figure~\ref{rolesBrainNets} shows the distribution of the roles obtained from all the analyzed networks over the $z-pc$ parameter space.  Most of the nodes in the functional brain networks  ($\sim 98\%$) can be classified as non-hubs (indicated by the gray area in Fig.~\ref{rolesBrainNets}-(b)), while only a minority of them are module hubs ($\sim2\%$).  Non-hubs nodes were classified as ultra-peripheral (R1, $10.33\%$) having all their links within their own modules; peripherals (R2, $73.49\%$) with most links within their modules; or non hub-connectors (R3, $13.67\%$) with half of their links to other modules. This distribution of roles strongly contrasts with that obtained from random configurations (results not show) where most nodes have their links homogeneously distributed among all modules (R4 and R7). 

The anatomical distribution of the parameters  $z$ and $pc$ is depicted in Figure~\ref{rolesBrainDistrib}. Interestingly, this representation shows that the wiring structure of the brain has a non-homogeneous organization in terms of the $z-pc$ parameters distribution. Examples of the different behaviours that can be observed are: \textit{i)} subcortical structures (indicated by the orange arrow) display relatively high values for both $z$-score and $pc$ parameters, indicating a dense inter- and intra-modular connectivity ; \textit{ii)} nodes belonging to brain areas associated to the primary visual system (pointed by the red arrow) have a scatter connectivity, yielding low values for both $pc$ and $z$ parameters; \textit{iii)} precuneus and cyngular gyrus areas (indicated by the yellow arrow) have a dense intra-modular connectivity (high values of $z$ ) but few links to other modules (low values of $pc$);  \textit{iv)}  frontal areas and some visual regions related to associative functions (cian arrow)  present more connections to other modules, which is reflected in their low values of $z$ and relatively high values of $pc$.

\section{Conclusion}
In conclusion, here we address a fundamental problem in brain networks research: whether the spontaneous brain behavior relies on the coordination (integration) of a complex mosaic of functional brain modules (segregation).  By using a random walk-based method we have identified a non-random modular structure of functional brain networks. In contrast to current approaches~\cite{Cordes2001, Salvador2005a, Achard2006}, our procedure requires neither of signal averaging in predefined brain areas, nor the definition of seed regions, nor subjective thresholds to assess the connectivities. To our knowledge, this work provides the first evidence of a modular architecture in functional human brain networks at a voxel level. 

The modularity analysis of large-scale brain networks unveiled a modular structure in the functional connectivity. Although a one-to-one assignment of anatomical brain regions to each detected module is difficult to define, results reveal a strong correlation between the spatial distribution of the modules and some well-known functional systems of the brain, including some of the frequently reported circuits underlying the functional activity at rest~\cite{Cordes2001}.  It is worth to notice that, although the functional brain connectivity is strongly shaped by the underlying anatomical wiring (e.g. by the white matter pathways), future studied are needed to clearly examine the interplay between the structural substrate and the modular connectivity  inferred from brain dynamics~\cite{Honey09}.

Our findings are in full agreement with previous studies about the structure of human brain networks. First, we have confirmed the degree distribution presents a power-law behavior over a wide range of scales, implying that there are a small number of regions with a large number of connections. We also found that brain connectivity shows a degree of clustering that is one order of magnitude higher than that of the equivalent random networks while keeping similar efficiency values, suggesting that spontaneous brain behavior involves an optimized (in a SW sense) functional integration of distant brain regions~\cite{Salvador2005a, Achard2006, Eguiluz2005}.  Further, the intrinsic non-random modular structure suggested by the high values of the clustering index of brain networks was confirmed by a high degree of modularity obtained for the ensemble of subjects.

Although the mechanisms by which modularity emerges in complex networks are not well understood,  it is widely believed that the modular structure of complex networks plays a critical role in their functionality~\cite{Variano2004, Guimera2005}. Functional brain modules can be related to a local -segregate- information processing while inter-modular connections allows the integration of distant anatomo/functional brain regions~\cite{Frackowiak1997}.  On the other hand, the SW and scale-free characteristics of brain webs provide an optimal organization for the stability, robustnes, and transfer of information in the brain~\cite{Achard2006, Achard2007, Bassett2006}. The modular structure constitutes therefore an attractive model for the brain organization as it supports the coexistence of a functional segregation of distant specialized areas and their integration during spontaneous brain activity~\cite{Tononi1998, Sporns2000}.  Although the study of anatomical brain networks is a current subject of research, we suggest that a modular description might provide new insights into the understanding of human brain connectivity during pathological or cognitive states.

\acknowledgments
This work was supported by the EU-GABA contract no. 043309 (NEST) and CIMA-UTE projects.


\end{document}